# ENHANCED CONTROL OF EXCIMER LASER PULSE TIMING USING TUNABLE ADDITIVE NOISE

ROBERT MINGESZ, ANGELA BARNA, ZOLTAN GINGL, and JANOS MELLAR

*Department of Technical Informatics, University of Szeged*
*Árpád tér 2., 6720 Szeged, Hungary*

Recently we have shown a system developed to precisely control the laser pulse timing of excimer lasers [1]. The electronic circuit based on an embedded microcontroller and utilized the natural jitter noise of the laser pulse generation to improve the long term regulation of the delay of the laser related to an external trigger pulse. Based on our results we have developed an improved system that uses additional, programmable time delay units to tune the noise source to further enhance performance and allows reduction of complexity in the same time. A mixed-signal microcontroller generates a randomly dithered delay of the pulse generation moment to enhance the resolution and also runs a dedicated algorithm to optimize regulation. The compact, flexible hardware supports further enhancements; the signal processing algorithm can be replaced even by in-system reprogramming. Optimized processing and the relaxed hardware requirements may also support low-power operation, wireless communication, therefore the application possibilities may be extended to many other disciplines.

## 1. Introduction

Precise time-synchronization of excitations, events is very important in experimental research in many disciplines including high-energy physics, photosynthesis research, chemical relaxation time measurements, technical applications of time-to-digital conversion and many more [2, 3].

In systems, where the trigger signal causes a delayed event at the output, regulation of the delay is required to maintain precise timing if the system's properties are time dependent, therefore the delay would change in time. A good example is the hydrogen thyratron that is widely used in excimer, carbon dioxide and copper vapour pulsed gas lasers, fast kickers in high-energy processes. In such applications it is important to keep the uncertainty and precision of synchronization below 1-2ns. The switching time – often called anode delay – of the thyratrons typically fall in the range of 100ps to 1ns, however this time is not constant. It has a slow, more or less deterministic, slowly varying component mainly due to temperature changes and an additive random jitter component associated by the gas discharge in the thyratron. Since the latter component is unpredictable, it cannot be compensated by active control; however the slow changes can be eliminated by inserting a tunable delay element between the start pulse and the input of the thyratron [1].

In the following we present a brief overview of the regulation and show our improved solution to the problem for internally and externally triggered cases.

## 2. Delay controller hardware

The principle of controlling the delay of the laser is illustrated on Fig.1. A programmable delay can be added to the laser delay to keep the time between the trigger and laser pulse at the desired value. The actual delay value must be measured in order to calculate the programmable delay value; note that the laser delay can be calculated as well, since the programmed delay is known. The time delays can be measured by various time-to-digital converters [4, 5], some of them are based on high speed counters running at a certain update rate [6].

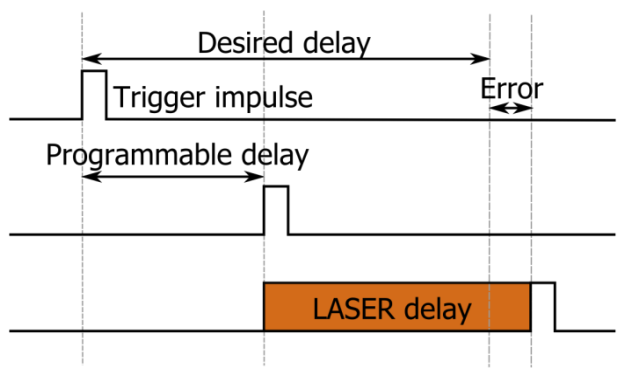

Fig. 1. The rising edge of the input pulse triggers a programmable delay element and this will start the laser. The delay of the whole system is the sum of the programmable delay and the laser inherent delay, thus the desired delay can be realized by compensating the varying delay of the laser.

In our original design [1] we have used integrated programmable delay lines to adjust the total delay of the system and a five-window digital window comparator to estimate the actual delay. The main drawback of this approach is the seriously limited range of the time-to-digital conversion, while the absence of internal triggering and the limitations of the programmable delay circuits reduce the performance of the system considerably as well. We have designed a new, compact, flexible microcontroller based hardware to perform all measurement and control operations. A C8051F120 microcontroller has a built in programmable counter array (PCA) that consists of six independent 16-bit compare/capture modules [7]. The main counter is running at 100MHz, and its actual value can be captured on the rising edge at the input of a module, therefore these modules can be used as time-to-digital converters with the resolution of 10ns, which is the system clock period. Note that this timebase if significantly more accurate than the timing defined by the delay line used in Ref [1].

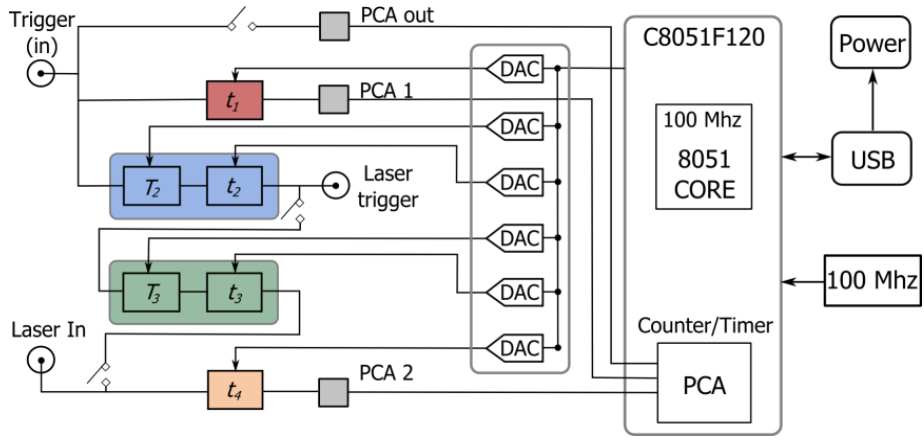

Fig. 2. Block diagram of the controller hardware. $t_1$, $t_2$, $t_3$ and $t_4$ are fine delay elements with range of 50ns, while $T_2$ and $T_3$ are coarse delay element to allow delays of 2500ns. The additional timing and control is provided by the embedded microcontroller.

The system contains four programmable delay units for different purposes:
- $t_1$ has a maximum range of 50ns, and it is used to add finely tuned delay for the start pulse before time-to-digital conversion by PCA module 1 (PCA1). It also allows random dithering to enhance resolution [8, 9].
- $T_2$ and $t_2$ are cascaded coarse and fine delay units with 2500ns and 50ns range to serve as a programmable delay to get the desired delay. These two units are required to have high enough resolution.
- $T_3$ and $t_3$ are cascaded coarse and fine delay units with 2500ns and 50ns range, and can be used to simulate the laser delay even without using the laser. They are useful to test and setup the unit or they can be used as general purpose delay elements.
- $t_4$ is a delay unit with range of 50ns and used to time shift the detection point of the laser pulse before time-to-digital conversion by PCA module 2 (PCA2). It can also be used to add random dither to enhance the resolution.

All of these programmable delay elements are based on a very simple principle [10]. The digital input signal is buffered and the output of this buffer charges a capacitor via a series resistor. The voltage $U_c$ at the capacitor approximates the supply voltage $U_s$ exponentially:

$$U_c(t) = U_s(1 - \exp(-t/RC)), \qquad (1)$$

where $R$ and $C$ is the value of the resistor and capacitor, respectively.

This voltage if fed into the noninverting input of a high-speed comparator (ADCMP603), whose inverting input voltage is set by a digital-to-analog converter (DAC). The output of the comparator will be switched when the capacitor will be charged to a voltage equal to the voltage at the comparator's inverting input. This means that the output is delayed depending on the output voltage of the DAC $U_{DAC}$:

$$t_{delay} = -RC \cdot \ln\left(1 - \frac{U_{DAC}}{U_s}\right). \qquad (2)$$

Note that the buffer and comparator have their own additive delays of a few nanoseconds, marked with circled numbers 1 and 2 on Fig.3.

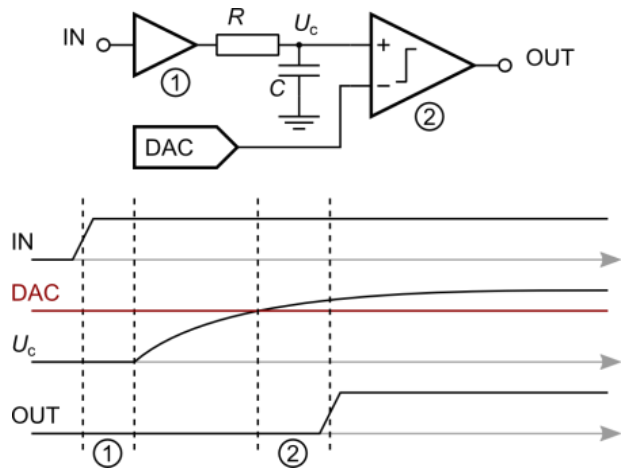

Fig. 3. Programmable delay generator. When the input signal goes from low to high state, the voltage $U_c$ at the output approximates the supply voltage exponentially. The digital-to-analog converter (DAC) is used to set the comparator level, thus the outputs signal is delayed depending on this level. Note that the input buffer and the comparator both have their own delay of a few nanoseconds labeled by 1 and 2 in circles and marked in the time diagram by dashed lines.

The six delay elements are programmed with a single integrated circuit containing six 12-bit DACs. The microcontroller can set all DAC output voltages and associated time delays and have a USB connection for optional supervised operation by a host computer.

The resolution of the 2500ns full scale range coarse delay units is 1ns with measured jitter of 90ps RMS, while the 50ns full scale range fine delay unit has resolution of 15ps and 10ps RMS jitter.

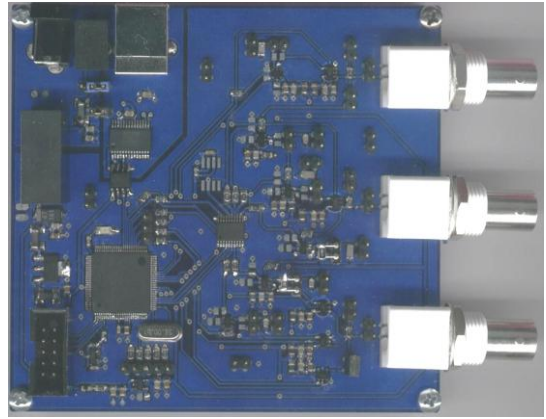

Fig. 4. Photo of the printed circuit board.

### 3. Delay control operation

Two different triggering modes can be realized depending on the source of the start impulse that triggers the laser. If the system is controlled by an external device, the start signal will be asynchronous to the system clock of the controller hardware. The controller can also be used to start the laser and other external devices; in this case the trigger impulse is synchronized to the 100MHz system clock.

#### 3.1. *Internally triggered operation*

As it was mentioned above, the laser can be triggered by the controller hardware, when the start pulse has a fixed delay from the preceding system clock toggle. The $T_2+t_2$ programmed delay and the laser delay are added to set the position of the laser pulse. The laser pulse triggers the $t_4$ delay element whose output will be captured by the PCA2 module. Fig.5. shows that this time instant can be synchronized to a system clock toggle point by tuning $t_4$. This means that the small jitter of the laser and the additional electronics will cause random flipping between two neighboring captured PCA2 values. This way the random jitter of the laser can be used to enhance the 10ns resolution by averaging the results obtained in a repetitive operation. It can also be desirable to add some jitter noise by setting $t_4$ randomly to optimize the effectiveness of the dither. Note that this delay does not affect the position of the laser output since it is only used to shift the counter capture signal. The desired delay is not limited to 10ns multiples, only the 15ps resolution of the fine delay unit limits the positioning.

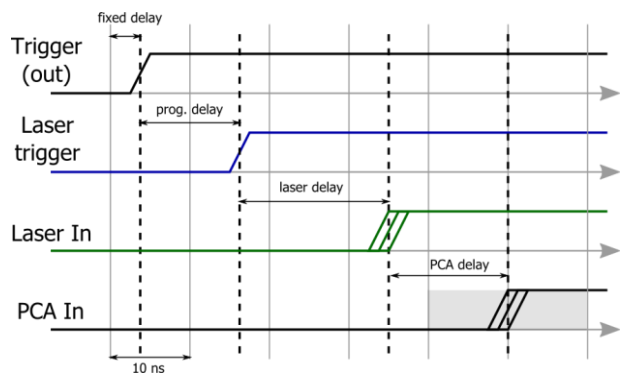

Fig. 5. Time diagram of the control signals. The trigger out signal is generated by the controller hardware therefore it is locked to the system clock which has a period of 10ns. The PCA is used to measure the resulting delay.

#### 3.2. *Externally triggered operation*

In many cases the laser start signal comes from an external device and must be synchronized to an external event. This means that neither the start pulse nor the laser output pulse is synchronous to the system clock. Therefore it is not any more possible to use the laser jitter alone as a dither signal to enhance resolution, since the capture signal cannot be moved to a system clock toggle point. In this case both the laser start signal and the laser output signal must be randomly dithered before time-to-digital conversion by PCA1 and PCA2 capture latches to allow resolution enhancement for repetitive operation. The uniformly distributed 10ns wide dither signal can be easily generated by programming the $t_1$ and $t_4$ delay elements with properly chosen pseudorandom numbers for the corresponding DACs. The time diagram of the signals is depicted on Fig.6.

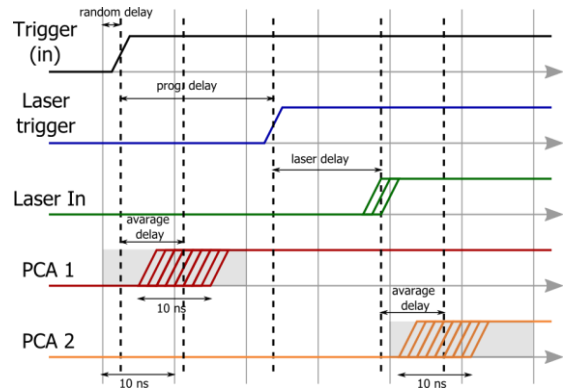

Fig. 6. Time diagram of the control signals. The trigger signal is generated externally, therefore it is not synchronized to the system clock therefore two PCA modules are used to measure the resulting delay. The 10ns wide additive dither on both PCA1 and PCA2 signals is used to enhance the 10ns resolution.

### 3.3. *Control algorithm*

We have implemented a rather simple and scalable control algorithm to keep the overall delay at the desired level even if the laser delay is changing in time. Note that the flexible hardware allows many solutions for different situations. The algorithm works both for the internally and externally triggered cases; the main difference is the method of measuring the delay as it was described above.

The flowchart and time diagram of the control principle can be seen on Fig.7. If the laser output is measured at the $k$-th 10ns window, i.e. the PCA2 module captures the value $k$, we can determine the laser delay by subtracting the actual programmed delay values. We can do an adaptive averaging on this fluctuating value to enhance resolution and to get a more precise result. The number of averages is determined during operation; we start with a value of 1 and increase this number if no significant change can be observed in the laser delay. The change detection threshold and maximum number of averages depend on many factors; they must be tuned to their optimum value in the real environment.

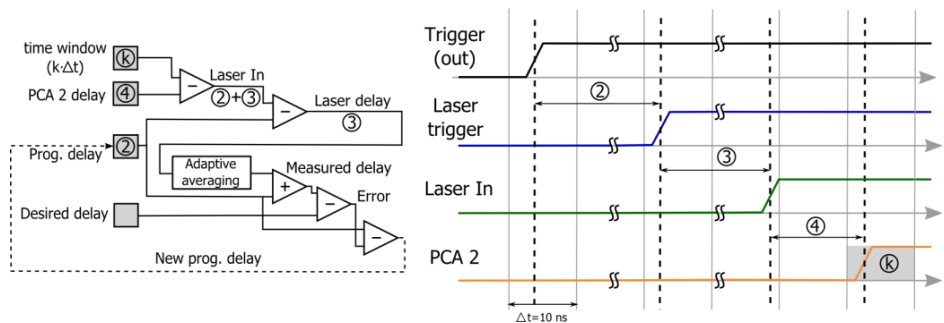

Fig. 7. Control algorithm and timing diagram of the signals. The PCA2 module is used to measure the whole delay. Since it has a resolution of 10ns, adaptive averaging is applied to enhance the resolution for repetitive signals.

### 3.4. *Performance of the control operation*

We have tested the performance of our control algorithm by numerical simulations using some typical cases and parameter values. The laser jitter was set to 1ns RMS for the simulation of the laser delay step and ramp responses. Fig.8. shows the result of the step response simulation. It can be seen, that due to the adaptive averaging a few laser shots are required to recover from an abrupt change in the laser delay. On the other hand, the algorithm only slightly increases the overall jitter as can be seen by comparing the delay plot with and without jitter.

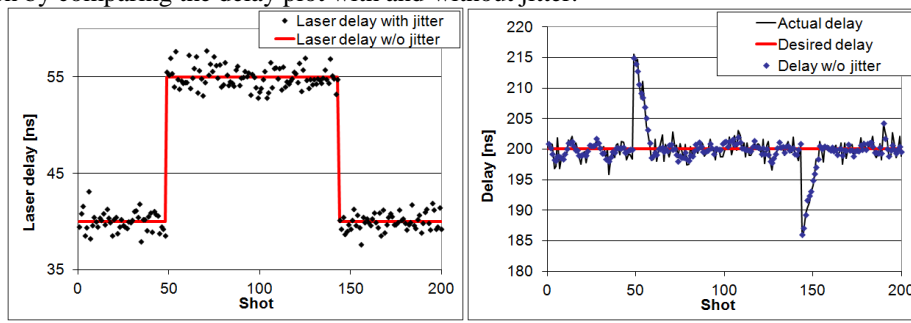

Fig. 8. Step response of the control algorithm. The laser delay was abruptly changed from 40ns to 55ns (left plot) while the desired delay was kept at 200ns. The simulated jitter noise of the laser was 1ns RMS. The signals are plotted both with and without this jitter.

Fig.9. illustrates the more natural ramp response simulation. During ramping the actual delay value is somewhat higher than the desired value, but is it still well below the inherent jitter of the laser. Ramping with higher slope may require different adaptive averaging parameters to have similar performance; however this situation is rather unlikely.

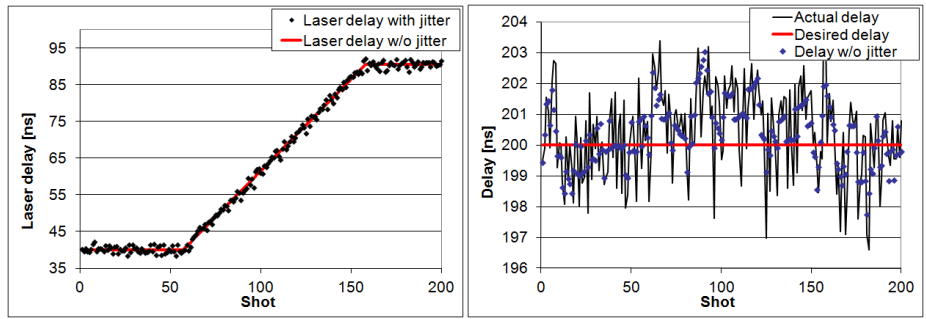

Fig. 9. Ramp response of the control algorithm. The laser delay was elevated changed from 40ns to 85ns (left plot) while the desired delay was kept at 200ns. The simulated jitter noise of the laser was 1ns RMS. The signals are plotted both with and without this jitter.

Note that all simulations show regulation precision close to 1ns although the resolution of the PCA timers is 10ns.

## 4. Conclusion

We have shown a rather simple solution to control laser pulse timing using the inherent jitter and additive random dither. We have designed and realized a microcontroller based hardware that has several programmable delay elements and timing units to support control and resolution enhancement by utilizing the different random noise sources. The device can be operated in a stand-alone mode when the embedded processor runs the control algorithm but can also be controlled by a host computer via the USB interface.

Compared to our former solution [1] – where we could only use the unpredictable jitter of the laser and electronic components while we were focusing on the fundamental stochastic resonance like characteristics of the system – here we have arrived at a more flexible approach that allows internal or external triggering; random dithering of both the trigger and laser output signals to improve time resolution; and a more simple and efficient control algorithm.

We also aimed to make the working principle versatile, and our low-cost device may also be used in many interdisciplinary applications where sub-ns time resolution and accurate timing control is required with high flexibility. In such applications, the main advantage is that even a single-chip mixed-signal microcontroller can be used with only a few passive external components. There are many microcontrollers on the market with integrated comparators, digital-to-analog converters and timing units, therefore the delay lines and time-to-digital conversion can all be easily implemented, no external circuits or costly application-specific circuits (ASICs) are needed. This allows various low cost ways to utilize the principle of using noise to enhance performance of the time control or just the time-to-digital conversion in many different applications. The USB interface can also be replaced by wireless communication and the system's power consumption can be optimized to support battery-powered operation.

**Acknowledgements**

This work was supported by grant OTKA K69018, Hungarian Academy of Sciences and TÁMOP-4.2.1/B-09/1/KONV-2010-0005.